\documentclass[a4paper]{article}

\usepackage{INTERSPEECH2019}
\usepackage{framed}

\title{Applying Speech Tempo-Derived Features, BoAW and Fisher Vectors\\to Detect Elderly Emotion and Speech in Surgical Masks}
\name{G\'abor Gosztolya$^{1,2}$, L\'aszl\'o T\'oth$^2$}
\address{
  $^1$MTA-SZTE Research Group on Artificial Intelligence, Szeged, Hungary\\
  $^2$University of Szeged, Institute of Informatics, Szeged, Hungary}
\email{\{ ggabor, tothl \} @ inf.u-szeged.hu}

\begin{document}

\maketitle
\begin{abstract}
The 2020 INTERSPEECH Computational Paralinguistics Challenge
(ComParE) consists of three Sub-Challenges, where the tasks are to
identify the level of arousal and valence of elderly speakers,
determine whether the actual speaker wearing a surgical mask, and
estimate the actual breathing of the speaker. In our contribution to
the Challenge, we focus on the Elderly Emotion and the Mask
sub-challenges. Besides utilizing standard or close-to-standard
features such as ComParE functionals, Bag-of-Audio-Words and Fisher
vectors, we exploit that emotion is related to the velocity of
speech (i.e. speech rate). To utilize this, we perform phone-level
recognition using an ASR system, and extract features from the
output such as articulation tempo, speech tempo, and various
attributes measuring the amount of pauses. We also hypothesize that
wearing a surgical mask makes the speaker feel uneasy, leading to a
slower speech rate and more hesitations; hence, we experiment with
the same features in the Mask sub-challenge as well. Although this
theory was not justified by the experimental results on the Mask
Sub-Challenge, in the Elderly Emotion Sub-Challenge we got
significantly improved arousal and valence values with this feature
type both on the development set and in cross-validation.
\end{abstract}

\section{Introduction}

Computational paralinguistics, a subfield of speech technology,
deals with extracting, locating and identifying various phenomena
being present in human speech. In contrast with Automatic Speech
Recognition (ASR), where most of such information is considered
secondary behind the phonetic content of the speech signal (i.e. the
phonetic or word-level transcription), computational paralinguistics
focuses on the huge variety of information related to the physical
and mental state of the speaker, usually ignoring the actual words
uttered. The Interspeech Computational Paralinguistics Challenge
(ComParE), held regularly at the Interspeech conference over a
decade now, focuses on the automatic identification of this
`paralinguistic' (that is, `beyond linguistic') aspect of human
speech. The open tasks presented over the years covered dozens of
different human speech aspects, ranging from emotion
detection~\cite{compare2009} through determining speaker age and
gender~\cite{compare2010}, estimating blood alcohol
level~\cite{compare2011} and identifying specific disorders which
affect the speech of the subject (e.g. autism~\cite{compare2013} and
Parkinson's Disease~\cite{compare2015}).

During the history of the ComParE Challenge, we can see two main
types of solutions for the various tasks. The first employs {\it
general} techniques, that might be applied for a wide range of
problems. Perhaps the most straightforward such technique is the
6373-sized `ComParE functionals' attribute set, which uses means,
standard deviations, percentile statistics (e.g. 1st, 99th), peak
detection etc. to form utterance-level attributes from certain
frame-level feature vectors. This feature set was also developed
over the years, taking its final form in 2013 (for the details, see
the work of Schuller et al.~\cite{compare2013}). Another such
approach is the Bag-of-Audio-Words (or
BoAW,~\cite{pancoast2012bagofaudiowords,pokorny2015detection,schmitt2016attheborder})
method, which first clusters the input frame-level feature vectors,
and then assigns each frame of each utterance into one of these
clusters and uses a statistics of these clusters to construct an
utterance-level feature vector. This technique is incorporated into
the Challenge baselines since 2017~\cite{compare2017}. Some other
feature extraction methods, albeit being of a general nature, were
only employed by certain participants so far, such as Fisher
vectors~\cite{kaya2015fishervectors,gosztolya2019usingfisher,wu2019thedkulenovo}.

The second type of approaches seek to employ {\it task-specific}
techniques. Clearly, one might expect that a solution designed and
fine-tuned for the actual problem at hand allows higher performance,
leading to better accuracy scores; on the other hand, they take more
effort to develop. For example, Gr\`ezes et al. calculated the ratio
of speaker overlap to aid conflict intensity
estimation~\cite{grezes2013letmefinish}; Montaci\'e and Caraty
detected temporal events (e.g. speech onset latency, event starting
time-codes, pause and phone segments) to detect cognitive
load~\cite{montacie2014highlevel}, several authors extracted phone
posterior-based attributes to determine the degree of nativeness or
the native language of the
speaker~\cite{montacie2015phraseaccentuation,shivakumar2016multimodalfusion,gosztolya2016determining},
while Huckvale and Beke developed specific spectral-based attributes
to detect whether the speaker has a
cold~\cite{huckvale2017itsoundslike}. Of course, some kind of fusion
of the general and the task-specific attributes might also prove to
be beneficial.

In our actual contribution to the ComParE 2020
Challenge~\cite{compare2020}, we apply specific task-dependent
attributes. It is well-known that the mental state of the subject
affects several prosodic and temporal properties of his speech;
specifically, emotion is strongly related to speech
tempo~\cite{arnfield1995emotionalstress,braun2007speakingtempo}, and
it affects the amount of hesitation as well. This means that, by our
hypothesis, calculating temporal parameters such as articulation
tempo (i.e. phones uttered per second), speech tempo and some
pause-related attributes, we might be able to estimate the emotional
state of the speaker. In other paralinguistic tasks, these
attributes might be indicators of different speaker states, e.g.
feeling uneasy when forced to speak in a surgical mask. Of course,
competitive performance is probably achieved via a combination of
predictions with those obtained by using standard approaches, such
as ComParE functionals or Bag-of-Audio-Words (BoAW).

Following the Challenge guidelines (see~\cite{compare2020}), we will
omit the description of the tasks, datasets and the method of
evaluation, and focus on the techniques we applied. Since the
Breathing sub-challenge is essentially a frame-level (or a {\it
few-frame-level}) task, which calls for entirely different
techniques, we focus on the remaining two sub-challenges in this
study: in the Mask Sub-Challenge (MSC), the task is to recognise
whether the speaker was recorded while wearing a surgical mask or
not, while in the Elderly Emotion Sub-Challenge (ESC) the task is to
determine the affective state of subjects aged 60 or over. While the
former is a binary classification task, in the Elderly Emotion
Sub-Challenge both arousal and valence have to be classified as low,
medium or high, therefore it essentially consists of two three-class
classification tasks. Classification performance is measured via the
Unweighted Average Recall (UAR) metric; for the Elderly Emotion
task, the UAR values corresponding to arousal and valence are
averaged out.

\section{Temporal Speech Features}

Next we describe the temporal speech features we extracted from the
utterances of the Elderly Emotion and Mask Sub-Challenges. We would
like to note that this attribute set was based on our previous works
focusing on detecting Mild Cognitive Impairment (MCI), Alzheimer's
Disease (AD) and schizophrenia (SCH) (see
e.g.~\cite{toth2018asrbased,gosztolya2019identifyingmild,gosztolya2018identifying}),
with some straightforward changes. That is, we removed the
calculation of the utterance length, as it was meaningless for the
short speech chunks provided for the particular sub-challenges. For
the list of our temporal parameters, see
Table~\ref{table_parameters}.

These speech parameters rely on the concept of {\it hesitations}.
The simpler form of pause or hesitation is that of a {\it silent
pause}: the absence of speech. However, {\it filled pauses} (sounds
like ``er'', ``um'' etc.) also indicate hesitations, and can take up
a significant amount of speech time. For example, T\'oth et al.
found that about 10\% of the hesitations in a Hungarian speech
database appear as filled pauses~\cite{toth2015automatic}. While the
simplest of our attribute, speech tempo corresponds to the average
number of phones found in one second of the utterance, in
articulation rate we take into account only those phones which are,
in fact, are not hesitations. The remaining attributes (i.e.
(3)-(6)) all describe the amount of hesitation within speech, but in
different ways. Furthermore, when we describe the amount of pauses,
we can take into account only silent pauses, only filled pauses, or
any of them; so the temporal parameters (3)-(6) can be calculated in
three variations, leading to a 14-sized attribute set.

To calculate these temporal parameters, first we performed speech
recognition; as we were interested in these specific parameters, we
decided to work only on the level of phones. While completely
discarding a word-level language model (even as simple as a
vocabulary) probably increases the number of errors in the ASR
output, notice that now we do not need to be able to accurately {\it
identify} the phones: all we need to do is to {\it count} them. We
need to identify only two phenomena: silences (including breath
intakes and sighs) and filled pauses. In this approach, we treated
filled pauses as a special `phoneme'\footnote{Yes, the quotes are
there for a reason, Mr. Moore~\cite{moore2019ontheuse}}; from our
previous experience, we also expected that silent and filled pauses
can be identified with a high accuracy.

Although the speech material of both the Elderly Emotion and the
Mask Sub-Challenges contained German speech, due to the absence of a
German speech corpus we trained our DNN acoustic models on Hungarian
speech samples. We were able to exploit this since both silent and
filled pauses appear to be quite language-independent and because
these two languages are quite similar on the phonetic level. We used
a roughly 44 hours subset of the BEA
corpus~\cite{neuberger2014development}, where the annotation
included several non-verbal acoustic cues such as breath intakes,
sighs, coughs, and most importantly, filled pauses. We used our
custom DNN implementation~\cite{toth2015phonerecognition}, and a
modified version of HTK~\cite{htkbook} for decoding, using a
(Hungarian) phone bi-gram as a language model.

\begin{table}
\begin{framed}
\begin{itemize}
\item[(1)]{{\bf Speech tempo}: the number of phones per second
(including hesitations).}
\item[(2)]{{\bf Articulation rate}: the number of phones
per second during speech (excluding hesitations).}
\vspace{2mm} \hrule \vspace{1mm}
\item[(3)]{{\bf Pause occurrence rate}: divide the
total number of pauses by the number of phonemes in the utterance.}
\item[(4)]{{\bf Pause duration rate}: divide the
total duration of pauses by the length of the utterance.}
\item[(5)]{{\bf Pause frequency}: divide the
number of pause occurrences by the length of the utterance.}
\item[(6)]{{\bf Average pause duration}: divide the
total duration of pauses by the number of pauses.}
\end{itemize}
\end{framed} \caption{{\it The examined temporal speech parameters,
based on the work of Hoffmann et al.~\cite{hoffmann2010temporal} and
T\'oth et al.~\cite{toth2018asrbased}.}} \label{table_parameters}
\end{table}

\section{General Feature Extraction Methods}

Next, we briefly describe the three standard feature extraction
approaches we utilized in the ComParE 2020 Challenge.

\subsection{`ComParE functionals' Feature Set}

Firstly, we used the 6373 ComParE functionals (see
e.g.~\cite{compare2013}), extracted by using the openSMILE
tool~\cite{opensmile}. The feature set includes energy, spectral,
cepstral (MFCC) and voicing related frame-level attributes, from
which specific functionals (like the mean, standard deviation,
percentiles and peak statistics) are computed to provide
utterance-level feature values.

\subsection{Bag-of-Audio-Words Representation}

The BoAW approach also seeks to extract a fixed-length feature
vector from a varying-length
utterance~\cite{pancoast2012bagofaudiowords}. Its input is a set of
frame-level feature vectors such as MFCCs. In the first step,
clustering is performed on these vectors, the number of clusters
($N$) being a parameter of the method. The list of the resulting
cluster centroids will form the {\it codebook}. Next, each original
feature vector is replaced by a single index representing the
nearest entry in the codebook ({\it vector quantization}). Then the
feature vector for the given utterance is calculated by generating a
histogram of these indices, usually after some kind of normalization
(e.g. in L1 normalization we divide each cluster count by the number
of frames in the given utterance).

To calculate the BoAW representations, we utilized the OpenXBOW
package~\cite{schmitt2017openxbow}. We tested codebook sizes of $N =
32, 64, 128, 256, 512, 1024, 2048, 4096, 8192$ and $16384$. We
employed random sampling instead of kmeans++ clustering for codebook
generation~\cite{schmitt2016attheborder}, and employed 5 parallel
cluster assignments;
otherwise, our setup followed the ComParE 2020 baseline paper
(i.e.~\cite{compare2020}): we used the 65 ComParE frame-level
attributes as the input after standardization,
and a separate codebook was built for the first-order derivatives.

\subsection{Fisher Vector Representation}

The aim of the Fisher vector representation is to combine the
generative and discriminative machine learning approaches by
deriving a kernel from a generative model of the
data~\cite{jaakkola1999exploiting}. First we describe the original
version, developed for image representation; then we turn to the
application of Fisher vectors to audio.

The main concept of the Fisher Vector (FV) representation, adapted
to audio processing, is to take the frame-level feature vectors of
some corpus and model their distribution by a probability density
function $p(X|\Theta)$, $\Theta$ being the parameter vector of the
model. For example, when using Gaussian Mixture Models with a
diagonal covariance matrix, $\Theta$ will correspond to the priors,
and the mean and standard deviation vectors of the components. The
Fisher score describes $X$ by the gradient $G^X_{\Theta}$ of the
log-likelihood function, i.e.
\begin{equation}
G^X_{\Theta}=\frac{1}{T}\nabla_{\Theta}\log p(X|\Theta).
\end{equation}
\noindent This gradient function describes the direction in which
the model parameters (i.e. $\Theta$) should be modified to best fit
the data. 
The Fisher kernel between the frame-level feature vector sequences
(i.e. utterances) $X$ and $Y$ is then defined as
\begin{equation}
K(X,Y)=G^X_{\Theta}F^{-1}_{\Theta}G^Y_{\Theta},
\end{equation}
\noindent where $F_{\Theta}$ is the Fisher information matrix of
$p(X|\Theta)$, defined as
\begin{equation}
F_{\Theta}=E_X[\nabla_{\Theta}\log p(X|\Theta) \nabla_{\Theta}\log
p(X|\Theta)^T].
\end{equation}
\noindent Expressing $F_{\Theta}^{-1}$ as
$F_{\Theta}^{-1}=L_{\Theta}^T L_{\Theta}$, we get the Fisher vectors
as
\begin{equation}
{\cal
G}^X_{\Theta}=L_{\Theta}G^X_{\Theta}=L_{\Theta}\nabla_{\Theta}\log
p(X|\Theta).
\end{equation}

We used the open-source VLFeat library~\cite{vedaldi2015vlfeat} to
fit GMMs and to extract the FV representation; we fitted Gaussian
Mixture Models with $N = 2, 4, 8, 16, 32, 64$ and $128$ components.
As the input frame-level feature vectors, we here again employed the
65 ComParE frame-level attributes; following our previous
experiments
(e.g.~\cite{gosztolya2019usingfisher,gosztolya2020usingthefishervector}),
we also employed the first-order derivatives (i.e. the $\Delta$
values).

\section{The Mask Sub-Challenge}

Firstly, we present our experimental results on the Mask
Sub-Challenge. For classification, we employed SVM with a linear
kernel, using the libSVM implementation~\cite{libsvm}; the value of
$C$ was set in the range $10^{-5}, 10^{-4}, \ldots, 10^1$.
To combine the different approaches, following our previous works,
we decided to take the weighted mean of the posterior estimates; the
weights were set on the development set, with 0.05 increments.

\begin{table}[t]
\renewcommand{\arraystretch}{1.1}
\caption{Results for the Mask Sub-Challenge} \label{table_res_mask}
\centering
\begin{tabular}{l|cc}
{\bf Approach}          & {\bf Dev} & {\bf Test} \\
\hline\hline
Temporal parameters     &~50.6\%~ &~---~ \\ 
\hline
ComParE functionals     &~64.2\%~ &~---~ \\ 
Bag-of-Audio-Words      &~64.5\%~ &~---~ \\ 
Fisher Vectors          &~67.7\%~ &~---~ \\ 
\hline
ComParE + Temporal      &~64.2\%~ &~---~ \\ 
ComParE + BoAW          &~65.8\%~ &~---~ \\ 
ComParE + FV            &~68.1\%~ &~72.0\%~ \\ 
\hline
ComParE + BoAW + FV     &~68.5\%~ &~71.8\%~ \\
All four attribute sets &~68.6\%~ &~71.6\%~ \\ 
\hline
Best single method in~\cite{compare2020} (test) &~63.4\%~ &~70.8\%~ \\
ComParE 2020 baseline~\cite{compare2020} &~---~ &~71.8\%~ \\
\end{tabular}
\end{table}
Our results achieved can be seen in Table~\ref{table_res_mask}. By
using the `ComParE functionals' feature set we got a slightly better
UAR score (at least, on the development set) than what was reported
in the baseline paper (i.e. 62.6\%~\cite{compare2020}), but it is
probably due to the different SVM implementation used (i.e. libSVM
instead of scikit-learn); on the other hand, Bag-of-Audio-Words led
to a quite similar classification performance (i.e. 64.2\% vs.
64.5\%). Unfortunately, using the temporal parameters turned out to
be much less beneficial: the 50.6\% UAR score measured on the
development set is only slightly higher than what is achievable by
random guessing.

When combining the approaches, the temporal parameters were not
really useful either: fusing them with the `ComParE functionals'
predictions actually brought an insignificant improvement (0.01\%);
furthermore, the ComParE + BoAW combination was also only slightly
better than any individulat method (i.e. cca +1\%). Fusing the
ComParE functionals predictions with those of the Fisher vectors,
however, led to an efficient machine learning model, achieving an
UAR value of 68.1\% on the development set. Adding the
Bag-of-Audio-Words and the temporal feature sets to this combination
did not help the prediction significantly on the development set;
or, according to our submissions, they even decreased the UAR
values. While the ComParE + FV variation achieved a 72.0\% on the
test set, even slightly outperforming the official baseline score
(which was obtained by a combination of models based on their test
set performance), we got 71.6\% and 71.8\% in the other two cases.
This, in our opinion, indicates that the Fisher vector
representation approach was quite robust on the Mask Sub-Challenge;
on the other hand, the temporal attributes were not really useful.
Bag-of-Audio-Words, on the other hand, were found to be quite
sensitive to meta-parameters, and are, in general, less robust than
either ComParE functionals or Fisher
vectors~\cite{gosztolya2018generalutterancelevel}.

\section{The Elderly Emotion Sub-Challenge}
\begin{table*}[t]
\renewcommand{\arraystretch}{1.1}
\caption{Results for the Elderly Emotion Sub-Challenge}
\label{table_res_elderly} \centering
\begin{tabular}{l|ccc|ccc}
                        & \multicolumn{3}{c|}{\bf Arousal} & \multicolumn{3}{c}{\bf Valence} \\
\cline{2-4}\cline{5-7}
{\bf Approach}          & {\bf Dev.} & {\bf CV} & {\bf Test} & {\bf Dev.} & {\bf CV} & {\bf Test} \\
\hline\hline
Temporal parameters     &~41.8\%~ &~39.5\%~ &~---~ &~33.3\%~ &~34.9\%~ &~---~ \\ %
\hline
ComParE functionals     &~35.4\%~ &~38.4\%~ &~---~ &~39.1\%~ &~37.7\%~ &~---~ \\
BERT embeddings (A+V)   &~35.0\%~ &~41.0\%~ &~---~ &~49.1\%~ &~64.0\%~ &~---~ \\
Fisher Vectors          &~37.8\%~ &~44.1\%~ &~---~ &~---~ &~---~ &~---~ \\ %
\hline
ComParE + Temporal parameters   &~46.8\%~ &~40.5\%~ &~42.7\%~ &~43.7\%~ &~37.7\%~ &~---~ \\
ComParE + BERT          &~38.7\%~ &~42.8\%~ &~---~ &~51.6\%~ &~64.5\%~ &~---~ \\
%
ComParE + Fisher Vectors  &~37.8\%~ &~44.1\%~ &~---~ &~---~ &~---~ &~---~ \\
BERT + Temporal parameters  &~45.3\%~ &~44.7\%~ &~---~ &~49.9\%~ &~64.5\%~ &~47.8\%~ \\
\hline
All attributes          &~49.1\%~ &~52.8\%~ &~53.2\%~ &~---~ &~---~ &~32.4\%~ \\
\hline
ComParE 2020 baseline~\cite{compare2020} &~---~ &~---~ &~49.8\%~ &~---~ &~---~ &~49.0\%~ \\
\end{tabular}
\end{table*}

This Sub-Challenge was quite different than either the Mask
Sub-Challenge, or most sub-challenges in the past years. The reason
for this is that the organizers provided features based on the
transcription of the utterances; but since these make sense only for
larger utterances than the standard few-seconds-long chunks,
predictions have to be submitted for recordings being several
minutes long. Unfortunately, this also meant that the training,
development and test sets all consisted of 87-87 (albeit long)
utterances. Another, although minor difference was the presence of
the two subtasks (i.e. arousal and valence).

On the technical level, this affected some parts of our the
classification framework as well. We decided to discard the chunks
provided by the organizers, and we focused on the longer recordings
(which we reconstructed by simply merging the 5-second-long chunks).
To compensate for the significantly less examples, we used 10-fold
(speaker-independent) cross-validation for meta-parameter setting
instead of relying on the provided development set; test set
predictions were made with the same SVM models. On the other hand,
to meet the Challenge guidelines, we repeated all experiments with
the provided train-dev setup.

Similarly to the Mask Sub-Challenge, we used the same libSVM
implementation (with the same $C$ values tested). On the other hand,
as the distribution of the Low, Medium and High class labels was
somewhat imbalanced, we decided to opt for downsampling. Since
downsampling shrinks our already small training sets even further,
we decided to repeat SVM training 100 times for each training fold;
therefore, for each feature set and for each $C$ value, we trained
1000 models. Model fusion was done by simply taking the (unweighted)
mean of the predicted posterior values.

Our results can be seen in Table~\ref{table_res_elderly}. First,
notice that the CV and the development set-level UAR scores not
always display the some tendencies. In our opinion, this is due to
the extremely small-sized corpus: having only $3\times87$ utterances
carries the risk to be insufficient even to allow measuring the
classification performance reliably. (Of course, this comes from the
attempt to provide BERT embeddings as features, which make sense
only for larger utterances.) Unfortunately, this also means that
setting the meta-parameters of the different methods might prove to
be challenging, which actually coincides with our experience. During
our experiments, we found that optimal meta-parameters (and the
corresponding accuracy / UAR values) which we set on the classic
``training set + development set'' set-up differed greatly from
those set in ten-fold cross-validation.

Regarding the arousal values, all tested approaches were proven to
be useful even alone to achieve a competitive performance; however,
among them, the temporal parameters and the Fisher vectors seem to
be the most effective techniques. For valence, however, the
linguistic attributes (i.e. the BERT embeddings) seem to be
unmatched: no other method was able even to come close to the 49.1\%
(development set) and the 64.0\% (cross-validation) UAR scores. It
is logical, though, as the other approaches are all acoustic ones,
and therefore not really suitable to detect
valence~\cite{compare2020}.

\section{Conclusions}

For our contribution to the Interspeech 2020 Computational
Paralinguistics Challenge, first we experimented with features
derived from speech tempo. Our motivation was that emotion is
reported to be related articulation tempo (i.e. the number of phones
uttered per second), and it affects the amount of hesitation as
well. To this end, we employed ASR techniques and extracted
articulation rate, speech tempo and further 12 attributes describing
the amount of hesitation in the utterance. According to our
experimental results on the development set, this attribute set is
not really useful for detecting whether the speaker is wearing a
surgical mask, as the UAR score of 50.6\% attained is only slightly
above the change achievable via random guessing; on the other hand,
in the arousal subtask of the Elderly Emotion Sub-Challenge it led
to similar UAR values as the other methods described in the baseline
paper.

Besides these custom features, we also applied standard methods like
Bag-of-Audio-Words and Fisher vectors, and combined our predictions
with those got by using the standard `ComParE functionals' attribute
set and in the case of the Elderly Emotion Sub-Challenge, the
various BERT embeddings. These methods and their combinations proved
to be quite useful on the development sets, but for the Mask
Sub-Challenge we even managed to outperform the official baseline
score, which itself is also a combination of four approaches.

For the Elderly Emotion Sub-Challenge, we found the tested temporal
attributes helpful for the arousal task; for valence, however, they
were less useful. In general, the low number of training,
development and test instances made Elderly Emotion a particularly
challenging\footnote{No pun intended} task; in the end, we managed
to obtain a mean UAR score of 50.5\%, but this is probably not much
higher than what could be achievable by random guessing (and saving
the fifth submission to pair up the best arousal and valence
``predictions'').


\section{Acknowledgements}

This research was partially supported by grant
TUDFO/47138-1/2019-ITM of the Ministry for Innovation and
Technology, Hungary and by the National Research, Development and
Innovation Office of Hungary via contract NKFIH FK-124413. L. T\'oth
and G. Gosztolya were also funded by the J\'anos Bolyai Scholarship
of the Hungarian Academy of Sciences and by the Hungarian Ministry
of Innovation and Technology New National Excellence Program
\'UNKP-19-4.

\bibliographystyle{IEEEtran}

\bibliography{2020-interspeech-compare}


\end{document}